\documentclass[aps,prl,showpacs,twocolumn,amsmath,amssymb]{revtex4-1}
\usepackage{graphicx,bbm}

\newcommand{\be}{\begin{equation}}
\newcommand{\ee}{\end{equation}}

\newcommand{\un}[1]{{\underline{#1}}}
\newcommand{\bra}[1]{{\langle #1 \vert}}

\newcommand{\ket}[1]{{\vert #1 \rangle}}

\newcommand{\ii}{ {\rm i} }
\newcommand{\dd}{ {\rm d} }
\newcommand{\NN}{\mathbb{N}}
\newcommand{\ZZ}{\mathbb{Z}}

\newcommand{\CC}{\mathbb{C}}

\newcommand{\y}{{\rm y}}
\newcommand{\x}{{\rm x}}
\newcommand{\z}{{\rm z}}
\newcommand{\LL}{{\hat {\cal L}}}

\newcommand{\mm}[1]{{\mathbf{#1}}}

\def\tr{{\,{\rm tr}}}

\def\one{\mathbbm{1}}

\def\tr{\,{\rm tr}\,}
\def\ket#1{|#1\rangle}
\def\bra#1{\langle#1|}

\def\ii{{\rm i}}

\def\z{{\rm z}}

\def\etal#1{#1}

\def\tit#1{}

\begin{document}

\title{Exact Nonequilibrium Steady State of an Open Hubbard Chain}
 
\author{Toma\v z Prosen}
\affiliation{Department of Physics, Faculty of Mathematics and Physics, University of Ljubljana, Ljubljana, Slovenia}

\date{\today}

\begin{abstract}
We discuss current carrying non-equilibrium steady state of an open fermionic Hubbard chain that is strongly driven by markovian incoherent processes localized at the chain ends.
An explicit form of exact many-body density operator for any value of the coupling parameter is presented. The structure of a matrix product form of the solution is encoded in terms of a novel diagrammatic technique which should allow for generalization to other
integrable non-equillibrium models.
\end{abstract}

  
\maketitle

The single-band fermionic Hubbard model is the key paradigm of many-body quantum physics. In spite of being conceptually extremely simple, involving only coherent hopping (tunneling) and on-site electron-electron interaction, the model is believed to describe fundamental phenomena, in particular in two dimensional lattices where it is believed to be the model of superconductivity in cuprates. In one dimension (1D), the Hubbard Hamiltonian has been diagonalized by coordinate Bethe ansatz by Lieb and Wu \cite{lieb}, while later Shastry completed the toolbox of algebraic Bethe ansatz~\cite{shastry:88} by finding a non-trivial $R-$matrix satisfying the star-triangle equation. 
However, even in 1D, these existing (zero temperature or equilibrium) exact solutions \cite{book} seem to be useless for describing interesting physics far from equilibrium, either time-dependent \cite{enss:12}, or steady-state \cite{pz12}.

The Hubbard model is one of the prime candidates to model many fundamental and emergent equilibrium and non-equilibrium phenomena  in strongly correlated many body systems, many of which can nowadays be simulated in the laboratory \cite{bloch,schneider,esslinger,strohmaier} but still awaiting clear theoretical explanation. Among the key problems is the understanding of the breakdown of the 
Mott insulator by a strong bias or external field \cite{mottreview,oka1,oka2,eckstein,fabian,zala} and characterization of dynamics in terms of some (perhaps universal) non-equilibrium states \cite{silva}, in particular for systems with a lot of internal structure, such as integrable systems \cite{caux,kollar}. Even within the linear-response theory, the main question on precise conditions for quantum transport in 1D to be diffusive, ballistic, or anomalous is still open \cite{zp,transport1}, whereas the issue is somehwat better understood in the particular case of Heisenberg $XXZ$ spin $1/2$ chain due to recent numerical \cite{jacek,karrasch} and analytical \cite{prosen:11a,prosen:11b} advances.

One may describe a finite (say thermal, voltage, or chemical) bias on the system by means of a boundary driven quantum master equation where the incoherent processes, realized by the so-called jump operators, are localized at the system's boundaries. The fixed point of such a dynamical semigroup then gives the
many-body density operator in the non-equilibrium steady state (NESS). Recently, two techniques have been proposed to look for exact solutions of NESS in interacting spin chains, the main example being the $XXZ$ model. In the first approach \cite{prosen:11a,prosen:11b}, later referred to as the
{\em isolating defect operator} (IDO) method, the {\em matrix product operator} (MPO) form of NESS has been obtained by enforcing cancellation of all the terms for which a certain defect operator appears in the bulk (away from the boundaries).
This resulted in a peculiar homogeneous cubic algebra for the generating matrices of MPO. 
Later, this solution has been re-derived \cite{kps} in terms of a  {\em local operator `divergence'} (LOD) relation resulting in inhomogeneous quadratic algebra (in fact ${\mathfrak sl}_2$ and its $q$-deformation)
in close analogy to the treatment of classical stochastic exclusion processes \cite{schutz}. LOD has been in turn explained \cite{sutherland} as a consequence of infinitely-dimensional star-triangle equation at complex representation parameter \cite{pip13,pi13,iz13}. It remains unclear, however, if and how the two approaches are related.

In this Letter we write down an explicit form of NESS for the many-body boundary driven Lindblad equation for the fermi Hubbard chain. Identifying the key general aspects of the IDO technique the cancellation mechanism can be, in general, facilitated {\em locally} in terms of a particular graph, being trivial for the $XXZ$ model, but exhibiting quite a nontrivial structure in the present case. NESS density operator for an $n$-site chain is expressed in terms of an operator sum over all recurrent walks of length $n$ over the graph.
We outline a new, {\em  constructive} technique which has a potential of being  generalizable to other integrable non-equilibrium models.

We consider an $n$-site Hubbard chain, which may be conveniently formulated in terms of a spin $1/2$ ladder, i.e., using two sets of Pauli operators $\sigma^s_j,\tau_j^s$, $s\in{\cal J}:=\{+,-,0,\z\}$, $j\in\{1\ldots n\}$, $\sigma^0_j\equiv \one$, with the Hamiltonian
\be
H_n = \sum_{j=1}^{n-1} (\sigma_j^+ \sigma_{j + 1}^- + \tau_j^+ \tau_{j + 1}^- + {\rm H.c.}) + \frac{u}{4}\sum_{j=1}^{n}\sigma_j^{\rm z}\tau_j^{\rm z}
\label{eq:ladder}
\ee
with non-dimensional interaction strength $u$ (measured in units of hopping energy).
We seek a fixed point of the Liouville master equation~\cite{lin}
\be
\frac{\dd}{\dd t}\rho = \LL \rho := -\ii [H_n,\rho] + \sum_{l=1}^4 \Bigl(L_l \rho L^\dagger_l - \frac{1}{2} \{L^\dagger_l L_l,\rho\}\Bigr)
\label{eq:lind}
\ee
with boundary dissipative processes which incoherently create electrons at the left end and annihilate electrons at the right end, with the rate $\varepsilon$
\be
L_1 = \sqrt{\varepsilon} \sigma^+_1,\;
L_2 = \sqrt{\varepsilon} \tau^+_1,\;
L_3 = \sqrt{\varepsilon} \sigma^-_n,\;
L_4 = \sqrt{\varepsilon} \tau^-_n.
\label{eq:Ls}
\ee
The standard fermionic Hubbard Hamiltonian $H_n(u)=-\sum_{j,s} (c^\dagger_{s,j}c_{s,j+1}+{\rm H.c.}) +  u \sum_{j} (n_{\uparrow,j}-\frac{1}{2})(n_{\downarrow,j}-\frac{1}{2})$ is reconstructed via Jordan-Wigner transformation
 $c_{\uparrow,j}=P^{(\sigma)}_{j-1} \sigma_j^-$ and $c_{\downarrow,j}=P^{(\sigma)}_n P^{(\tau)}_{j-1} \tau_j^-$, $n_{s,j} := c^\dagger_{s,j}c_{s,j}$,
 where $P^{(\sigma)}_j:=\sigma_1^{\rm z} \cdots \sigma_j^{\rm z}$, $P^{(\tau)}_j:=\tau_1^{\rm z} \cdots \tau_j^{\rm z}$. It can be shown \cite{benenti:09} that, in the presence of local boundary dissipation, taking the jump operators as $c^\dagger_{\uparrow,1},c^\dagger_{\downarrow,1},c_{\uparrow,n},c_{\downarrow,n}$, the spin-ladder and fermionic models have equivalent NESSes.

\begin{figure}
 \centering	
\vspace{-1mm}
\includegraphics[width=0.9\columnwidth]{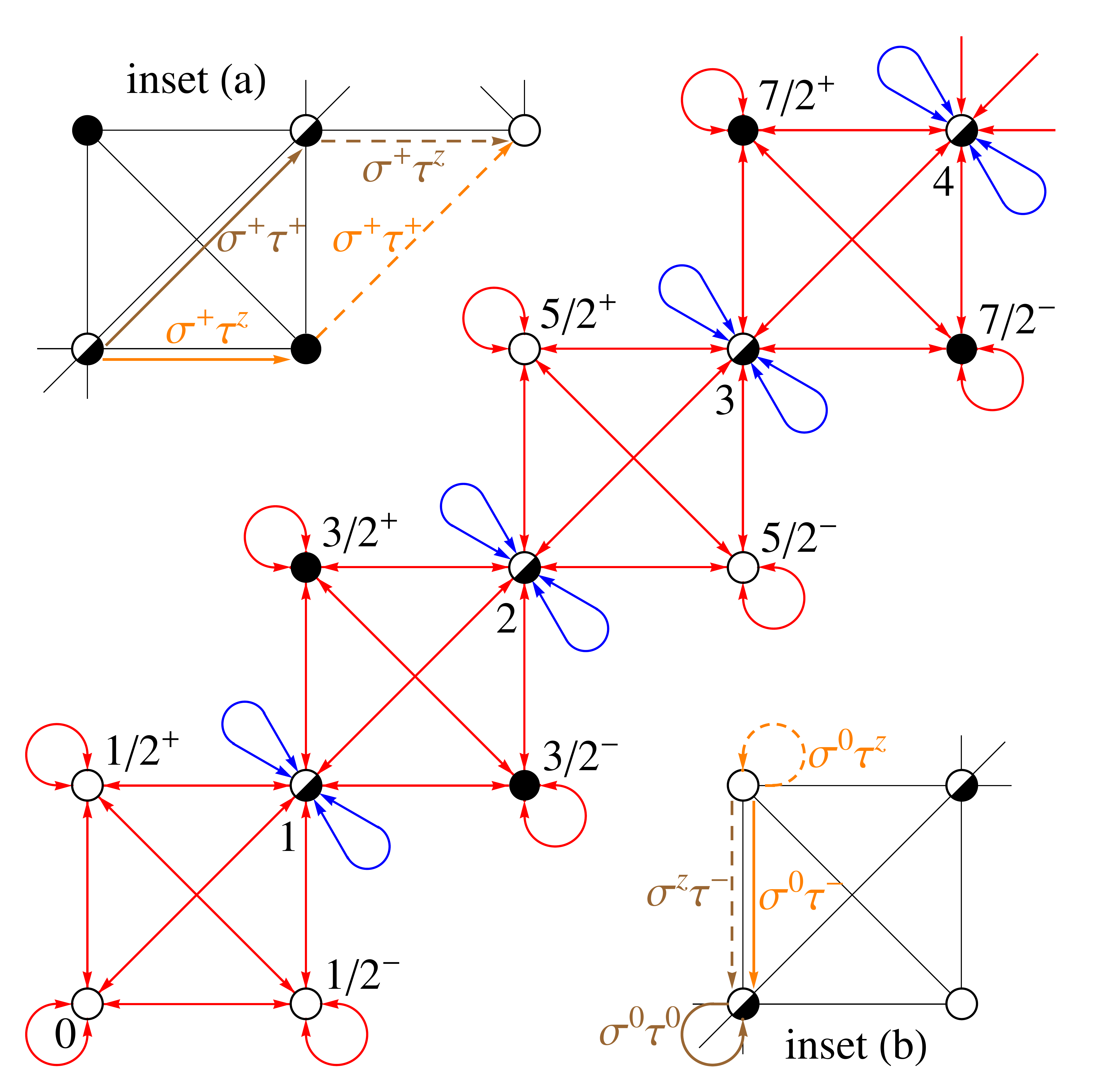}
\vspace{-1mm}
\caption{(Color online) A semi-infinite graph $G$ (structure repeating periodically beyond the upper-right corner) showing the {\em allowed} transitions 
for building up the MPO form of NESS for the Hubbard chain. 
Nodes in black, edges with multiplicity 1 in red, and edges with multiplicity 2 in blue.
Each edge $e$ is associated with a physical product-operator $\omega(e)=\sigma^{b^1}\tau^{b^2}$ 
where $b^\nu=0$ ($b^\nu=\z$) for edges connecting white (black) nodes, where
$\nu$ is that Cartesian component which does not change along such $e$ in the diagram.
Degenerate edges correspond to operators $\sigma^0 \tau^0$ ($\mu=+1$) and $\sigma^\z\tau^\z$ ($\mu=-1$). Insets indicate all possible terms (two in each, orange and brown) for two examples of $[h,\omega(e)\otimes\omega(f)]$,
namely $[h,\sigma^+\tau^+\otimes\sigma^+\tau^0]$ (a), and $[h,\sigma^0\tau^-\otimes\sigma^0\tau^0]$ (b). Full arrows denote valid edge factors, while dashed arrows correspond to {\em defect} operators.}
\label{fig:HD}
\end{figure}

The main result of this Letter is the following:

\noindent
{\bf Theorem:} A unique \cite{note} unnormalized NESS density operator of the boundary driven Hubbard chain (\ref{eq:ladder}-\ref{eq:Ls}) reads
\be
\LL\rho_\infty = 0,\quad \rho_\infty = S_n S_n^\dagger
\label{eq:cholesky}
\ee
where
\be
S_n=\sum_{\un{e} \in {\cal W}_n(0,0)} a_{e_1} a_{e_2} \cdots a_{e_n} \prod_{j=1}^n \sigma^{b^1(e_j)}_j \tau^{b^2(e_j)}_j.
\label{eq:wg}
\ee
${\cal W}_n(v,r)$ is a set of all $n-$step walks $\un{e}=(e_1,\ldots,e_n)$, $e_j$ being the corresponding directed edge at step $j$, 
starting at the node $v$ and ending at node $r$ of the directed graph $G$ depicted in Fig.~\ref{fig:HD}.
The set of nodes ${\cal V}(G)$ is composed of: the origin $0$, the {\em diagonal} nodes $k$, and upper-, and lower-diagonal nodes
$(k-\frac{1}{2})^+$, and $(k-\frac{1}{2})^-$, for $k \in \NN$. Each node $v \in {\cal V}(G)$ can also be identified with a pair of
Cartesian components $v\equiv (v^1,v^2)$ in the corresponding planar diagram (Fig.~\ref{fig:HD}), namely $k \equiv (k,k)$, $(k-\frac{1}{2})^+ \equiv (k-1,k)$, $(k-\frac{1}{2})^- \equiv (k,k-1)$.
The set of directed edges ${\cal E}(G)$ contains vertical, horizontal, diagonal, skew-diagonal, and self-connections, as indicated in Fig.~\ref{fig:HD}, where only self-connections
of diagonal nodes are degenerate with multiplicity two. Edges may also be identified with triples $e\equiv (p(e),q(e);\mu(e))$, pointing from node $p(e)$ to $q(e)$ and having degeneracy label $\mu(e)$, where $\mu=1$ for all edges except diagonal self-connections $(k,k;\mu)$ where $\mu \in\{\pm 1\}$.

To each edge $e \in {\cal E}(G)$ we associate a unique operator $\sigma^{b^1(e)} \tau^{b^2(e)} \equiv \omega(e)$ over $\CC^2\otimes\CC^2$ 
via {\em index functions} $b^{1,2} : {\cal E}(G) \to \{+,-,0,\z\}$ defined as follows:
$b^\nu (e)=\pm$ if $q^\nu(e)-p^\nu(e)=\pm 1$, while for $q^\nu(e)=p^\nu(e)$, $b^\nu(e)=0$, if $e$ connects {\em white} nodes, and
$b^\nu(e)=\z$, if $e$ connects {\em black} nodes. 
For diagonal self-connections (on {\em black-and-white} nodes), the index functions are determined by the degeneracy index, 
$b^\nu(k,k;1)=0, b^\nu(k,k;-1)=\z$.
To each node $v$ we associate a scalar or spinor vector space ${\cal H}_v$, namely for diagonal nodes ${\cal H}_v \equiv \CC^2$ while for the other nodes
${\cal H}_0,{\cal H}_{(k-1/2)^\pm} \equiv \CC^1$. To each edge $e$ we then associate a linear map $a_e : {\cal H}_{q(e)} \to {\cal H}_{p(e)}$, namely (omitting the degeneracy label when trivial):
\begin{widetext}
\begin{eqnarray}
a_{(0,0;+1)} &=& 1, \quad a_{(0,0;-1)} = 0,  \;\quad a_{(0,1)} = \begin{pmatrix}2\ii \varepsilon & 0 \end{pmatrix},\quad a_{(1,0)} = 
\frac{1}{2} 
\begin{pmatrix}
\ii\varepsilon - u \cr 
-2
\end{pmatrix},
\;\,\quad a_{(0,1/2^\pm)} = \varepsilon,
\quad a_{(1/2^\pm,0)} = -\ii, \nonumber \\
a_{(k,(k+1/2)^\pm)} &=& \begin{pmatrix}
\varepsilon \cr 0\end{pmatrix},
\quad\;\;
a_{(k,(k-1/2)^\pm)} = \frac{1}{4}\begin{pmatrix}
-(-1)^k (\ii \varepsilon - k u)\varepsilon \cr
(-1)^{\lfloor\frac{k+1}{2}\rfloor}2\varepsilon
\end{pmatrix}, 
\qquad\qquad\quad\quad\;\; a_{((k-1/2)^\pm,(k-1/2)^\pm)} = (-1)^k \frac{1}{2}\ii\varepsilon,
\nonumber \\
a_{((k-1/2)^\pm,k)} &=& \begin{pmatrix}
\varepsilon & 0
\end{pmatrix}, \quad
a_{((k+1/2)^\pm,k)} = \frac{1}{4}\begin{pmatrix}
-(-1)^k(4\ii - k \varepsilon u) & (-1)^{\lfloor\frac{k+1}{2}\rfloor} 2\varepsilon
\end{pmatrix},
\quad
a_{((k-1/2)^\pm,(k-1/2)^\mp)} = -\ii \varepsilon,
\nonumber\\
a_{(k,k+1)} &=& (-1)^k 2\ii\varepsilon \begin{pmatrix}
1 & 0\cr
0 & 0
\end{pmatrix},\qquad
a_{(k+1,k)} = \frac{1}{4}\begin{pmatrix}
(-1)^k (2\ii - \frac{1}{2}k \varepsilon u)(\varepsilon+ \ii (k+1) u) & -(-1)^{\lfloor\frac{k+1}{2}\rfloor}(\varepsilon+\ii(k+1)u)\varepsilon,\cr
-(-1)^{\lfloor\frac{k}{2}\rfloor}(\ii k \varepsilon u + 4) & -2\ii\varepsilon
\end{pmatrix},
\nonumber\\
a_{(k,k;(-1)^k)} &=&
\frac{1}{4}
\begin{pmatrix}
(-1)^k (\ii k \varepsilon u + 4) & (-1)^{\lfloor\frac{k+1}{2}\rfloor} 2\ii \varepsilon \cr
0 & 0 
\end{pmatrix}, 
\qquad
a_{(k,k;-(-1)^k)} =
\frac{1}{4}
\begin{pmatrix}
(-1)^k (\varepsilon+\ii ku)\varepsilon & 0 \cr
(-1)^{\lfloor \frac{k+1}{2}\rfloor} 2\ii\varepsilon & 0
\end{pmatrix}.  \label{givea} \end{eqnarray}
\end{widetext}
{\bf Proof:} We start by noting that walking graph state expression (\ref{eq:wg}) can be cast in the MPO form
\be
S_n = \!\!\!\!\!\sum_{s_1,t_1\ldots s_n,t_n\in{\cal J}}\!\!\!\!\!\!\bra{0}\mm{A}_{s_1,t_1}\cdots \mm{A}_{s_n,t_n}\ket{0}\prod_{j=1}^n \sigma_j^{s_j}\tau_j^{t_j},
\label{MPO}
\ee
by introducing a set of $16$ infinitely dimensional operators over auxiliary Hilbert space ${\cal H}=\bigoplus_{v\in{\cal V}(G)} {\cal H}_v$
\be
\mm{A}_{s,t} = \bigoplus_{e\in{\cal E}(G)} \delta_{s,b^1(e)} \delta_{t,b^2(e)} a_e.
\ee
and $\ket{0}$ being the state with component $1$ in ${\cal H}_0$ and $0$ elsewhere.
Note that $\mm{A}_{\z,0}=\mm{A}_{0,\z}=0$.
In full analogy with the proof for the $XXZ$ model of Ref.~\cite{prosen:11b}, i.e., by observing local properties of the dissipative part of $\LL$ (\ref{eq:lind}),
one shows that $\LL(S_n S_n^\dagger) = 0$ is implied by the relation
\begin{eqnarray}
[H_n,S_n]=\ii \varepsilon\!\!\sum_{s\in\{0,+\}}\!\!\bigl(\sigma^\z \tau^s \otimes P^{0,s}_{n-1} &+ \sigma^s \tau^\z \otimes P^{s,0}_{n-1}  - \nonumber \\
- Q^{0,-s}_{n-1}\otimes \sigma^\z \tau^{-s} &- Q^{-s,0}_{n-1}\otimes \sigma^{-s}\tau^\z\bigr),\quad \label{eq:defrel}
\end{eqnarray}
introducing the operators $P^{s,t}_{n-1},Q^{s,t}_{n-1}$ over $\CC^{4^{n-1}}$
\begin{equation}
P^{s,t}_{n-1} = \frac{\tr_{\! 1}\{(\sigma_1^s\tau_1^t)^\dagger S_n\}}{\tr (\{\sigma^s\tau^t)^\dagger \sigma^s\tau^t\}},\;
Q^{s,t}_{n-1} = \frac{\tr_{\! n}\{(\sigma_n^s\tau_n^t)^\dagger S_n\}}{\tr\{(\sigma^s\tau^t)^\dagger \sigma^s\tau^t\}}
\end{equation}
where $\tr_{\! j}$ denotes the partial trace with respect to $4$-dimensional local space at site $j$. Note the Hilbert-Schmidt orthogonality of Pauli products $\sigma^s\tau^t$.
The main part of the proof is then to show Eq.~(\ref{eq:defrel}) for ansatz (\ref{eq:wg}) with the amplitudes (\ref{givea}).

In order to do this, we elaborate here on {\em local} IDO method with respect to the graph $G$.
Let us consider an arbitrary walk of length 2, i.e., a pair of subsequent edges $e,f\in {\cal E}(G)$, with $q(e)=p(f)$.
Writing an arbitrary Hubbard type Hamiltonian density on a pair of sites as 
$h=(\sigma^+\tau^0\otimes \sigma^-\tau^0 + \sigma^0\tau^+\otimes\sigma^0\tau^- + h.c.)+u_1 \sigma^\z \tau^\z \otimes \one_4 + u_2 \one_4\otimes \sigma^\z\tau^\z$, one finds the following general form of the local commutator of $h$ with a tensor product of two valid edge factors for a pair of consecutive edges ($2-$walks) $e,f\in{\cal E}(G)$, $q(e)=p(f)$

\begin{widetext}
\begin{eqnarray}
&&[h,\omega(e)\otimes \omega(f)] = \!\!\!\!\sum_{s,t \in{\cal J},e'\in{\cal E}(G)}^{p(e')=p(e),q(f)-q(e')=d(s,t)}\!\!\!\!\!\!\!\! X^{s,t}_{e,f}\,\omega(e')\otimes \sigma^s \tau^t  + \!\!\!\!\sum_{s,t \in{\cal J},f'\in{\cal E}(G)}^{q(f')=q(f),p(e)-p(f')=d(s,t)}\!\!\!\!\!\!\!\! Y^{s,t}_{e,f}\,\sigma^s \tau^t \otimes \omega(f'), \label{local}
\end{eqnarray}
for suitable c-number coefficients $X^{s,t}_{e,f}(u_1,u_2),Y^{s,t}_{e,f}(u_1,u_2)$.
We define a displacement vector associated with a pair of Pauli indices, namely $d(\pm)=\pm 1, d(0)=d(\z)=0$, and write $(d(s),d(t))\equiv d(s,t)$.
Eq.~(\ref{local}) has the following crucial property: Any tensor factor $\sigma^s \tau^t$ in the first (or second) sum on RHS of (\ref{local}) is (i) 
neither of the form $\omega(f')$ (or $\omega(e')$), for {\em any} edge $f'$ (or $e'$) which would complete the 2-walk $(e',f')$ to connect the same nodes as $(e,f)$, 
(ii) nor is the missing link $d(s,t)$ between $q(e')$ and $q(f)$ (or $p(e)$ and $p(f')$) provided by {\em any} edge of $G$ at all! We shall call such a factor {\em a defect operator}. See insets of Fig.~\ref{fig:HD} for a few examples.
Since the Hamiltonian is a sum of local terms the entire commutator $[H_n,S_n]$ written  in the tensor product expansion (like (\ref{eq:wg})) is composed of terms which correspond to $n$-walks over a defected graph with
exactly {\em one} defect operator. As the RHS of (\ref{eq:defrel}) has only boundary defects, in the first or last factor, all the terms with defects in the bulk should therefore identically vanish.
Picking any pair of nodes, $v,r\in{\cal V}(G)$, which can be connected with at least one $3-$walk, it is then sufficient that the following local conditions are satisfied
\begin{equation}
\sum_{(e,f,g)\in {\cal W}_3(v,r)} a_e a_f a_g  \tr\!\left\{ \left(\omega(e')\otimes \sigma^s \tau^t\otimes\omega(g')\right)^\dagger [H_3,\omega(e)\otimes\omega(f)\otimes\omega(g)]\right\}
 = 0,
 \label{bulk}
\end{equation}
for any pair of edges $e',g'\in{\cal E}(G)$ for which $p(e')=v,q(g')=r$, and any defect component $s,t\in{\cal J}$. Of course, for many combinations $(v,r,e',g',s,t)$ the above equation is trivial, i.e. always satisfied, e.g., when $\sigma^s\tau^t=\omega(f')$ for some 
valid edge $f'$ between $q(e')$ and $p(g')$.
The remaining equations which need to be checked are those for which the defect operator sits at the first $j=1$ or the last $j=n$ tensor factor.
Again, one can factor out sufficient local conditions, which can now be formulated on two sites, in terms of $2-$walks, namely
\begin{eqnarray}
&& \sum_{(e,f)\in{\cal W}_2(0,v)} a_e a_f \tr\!\left\{\left(\sigma^s\tau^t \otimes \omega(f')\right)^\dagger\left([H_2,\omega(e)\otimes\omega(f)]- \ii \varepsilon \hat{\cal P}(\omega(e))\otimes \omega(f)\right)\right\}=0, \label{eq:lbc} \\
&&\sum_{(e,f)\in{\cal W}_2(v,0)} a_e a_f \tr\!\left\{\left(\omega(e')\otimes\sigma^s\tau^t\right)^\dagger\left([H_2,\omega(e)\otimes\omega(f)]+ \ii \varepsilon \omega(e)\otimes \hat{\cal P}(\omega(f))\right)\right\}=0, \label{eq:rbc}
\end{eqnarray}
\end{widetext}
for all $e',f' \in{\cal E}(G)$, with $q(f')=v$, and $p(e')=v$. 
 $\hat{\cal P}$ is a map over $4\times 4$ matrices defined as $\hat{\cal P}(\rho) := \frac{1}{2}\sigma^\z\otimes \tr_{\!\sigma}(\rho) + \frac{1}{2}\tr_{\!\tau}(\rho)\otimes \tau^{\rm z}$ where $\tr_{\!\sigma}$ (or $\tr_{\!\tau}$) 
denotes the partial trace over $\sigma$ (or $\tau$) qubit. Now, the set of possible defect operators is quite limited, namely $(s,t)\in\{(0,\z),(\z,0),(+,\z),(\z,+)\}$ for the left boundary conditions (\ref{eq:lbc}), or to $(s,t)\in\{(0,\z),(\z,0),(-,\z),(\z,-)\}$ for the right boundary condition (\ref{eq:rbc}).

Summarizing, checking all the three-point conditions in the bulk (\ref{bulk}) and the two-point boundary conditions (\ref{eq:lbc},\ref{eq:rbc}) is {\em sufficient} for establishing validity of Eq.~(\ref{eq:defrel}) for {\em any} $n$.
Verification of (\ref{bulk}-\ref{eq:rbc}) has been implemented by means of a computer algebra program in Mathematica. Since the amplitudes (\ref{givea}) are at most quadratic in the node label $k$, modulated with periodicity $4$ of sign factors $(-1)^{k},(-1)^{\lfloor k/2\rfloor},(-1)^{\lfloor (k+1)/2\rfloor}$, it is enough to check recurrence relations (\ref{bulk}) for sufficiently large {\em finite} piece of $G$ (comfortably estimating, for $k\le 28$). Thus, all that is needed to prove our solution rigorously for any $n$ has been done in finitely many computer steps. In fact, what has been done in practice, at first, is that Eqs.~(\ref{bulk}-\ref{eq:rbc}) have been used to compute the amplitudes $a_e$ recursively, for increasing node labels $k$.
This procedure has nevertheless been quite tedious, and we are unable to express it in a short algorithmic form.

Before closing, let us make a few remarks on the properties of our solution:
(i) Similarly to the solution \cite{prosen:11b} of $XXZ$ model, (\ref{eq:cholesky}) is again a Cholesky decomposition of the many-body density operator. Indeed, in the eigenbasis of $\sigma_j^\z,\tau_j^\z$, the operator $S_n$ is an upper-triangular matrix.
(ii) $S_n$ is a polynomial of degree not more than $2n$ in dissipation $\varepsilon$, and polynomial of degree not more than $n$ in interaction strength $u$.
(iii) Ordering basis sets in the auxiliary space with respect to the increasing node index $k$, the matrices $\mm{A}_{s,t}$ (generating MPO (\ref{MPO}))  are block tridiagonal, with blocks of size $4$. In fact, the maximal Schmidt rank for the bipartition of $S_n$ in the Pauli basis is $4\lfloor n/2\rfloor $.
(iv) Efficient computation of local observables, like spin or charge densities, currents, etc., can again be facilitated with a concept of a transfer matrix (which is now a block tridiagonal matrix) introduced in Refs.~\cite{prosen:11a,prosen:11b}.
Details will be given elsewhere.
(v) Following the idea of Ref.~\cite{kps} and writing a Lax operator over ${\cal H}\otimes \CC^4$, as $\mm{L} = \sum_{s,t\in{\cal J}}\mm{A}_{s,t}\sigma^s\tau^t$, Eq.~(\ref{eq:defrel}) follows if another operator $\mm{B}=\sum_{s,t\in{\cal J}}\mm{B}_{s,t}\sigma^s\tau^t$ over  ${\cal H}\otimes \CC^4$ exists such that LOD relation would hold
\begin{equation}
[h,\mm{L}\otimes_{\rm p}\mm{L}] = \mm{B}\otimes_{\rm p}\mm{L}-\mm{L}\otimes_{\rm p} \mm{B},
\label{LOD}
\end{equation} 
where $h=H_2(u/2)$ is a symmetric translationally invariant Hamiltonian density, and $\otimes_{\rm p}$ denotes a tensor product with respect to the physical spaces $\CC^4$ and ordinary matrix product in ${\cal H}$. 
Knowing the operator $\mm{L}$ explicitly the equation (\ref{LOD}) is an overdetermined set of linear equations for matrix elements of $\mm{B}$ and a simple computer-algebraic calculation suggests existence of a non-trivial solution with simple $4\times 4$ block tridiagonal form of matrices $\mm{B}_{s,t}$. 
However, any possible relationship to $SO(4) \simeq SU(2)\times SU(2)/\ZZ_2$ symmetry of the Hubbard model and its Yang-Baxter algebra \cite{book} remains open.
(vi) It can be shown that the dissipative boundary conditions break the global symmetry of the open Hubbard chain to $SU(2)\times U(1)$, i.e., $\rho_\infty$ commutes with generators $S^\pm,S^\z$ and $\eta^\z$ of Ref.~\cite{book}.
We find $\tr S^s \rho_\infty=\tr \eta^\z \rho_\infty = 0$ and $\tr[((S^\x)^2+(S^\y)^2+(S^\z)^2)\rho_\infty]/\tr\rho_\infty = 3n/8$. (vii) We also observe empirically, for small $n$, that similarly to $XXZ$ model \cite{pip13,pi13}, $S_n(\varepsilon)$ has a commuting-transfer-matrix property with respect to dissipation parameter, i.e., $[S_n(\varepsilon),S_n(\eta)]=0, \forall \varepsilon,\eta\in\CC$.  The first derivative $Z=-\ii (\dd/\dd\varepsilon)S_n|_{\varepsilon=0}$ gives a quadratically extensive almost conserved operator, $[H,Z]= \sigma_1^\z + \tau_1^\z - \sigma_n^\z - \tau_n^\z$, which should be explored in studying high-temperature transport 
properties \cite{transport} of Hubbard chains. (viii) The same NESS (\ref{eq:cholesky}) applies to more general, chemically shifted Hubbard Hamiltonians 
$H + \mu_\uparrow N_\uparrow + \mu_\downarrow N_\downarrow$, with $N_\uparrow = \sum_j \frac{1}{2}(1-\sigma^\z_j)$, $N_\downarrow = \sum_j \frac{1}{2}(1-\tau^\z_j)$, as clearly all walks in (\ref{eq:wg}) conserve $N_s$, implying $[N_s,\rho_\infty]=0$.
Yet, $\rho_\infty$ corresponds to, on average, half-filled state with zero magnetization, due to symmetry of the driving (\ref{eq:Ls}).

In conclusion, the results presented here have, on one hand, the potential to be applied to some of the outstanding problems for the 1D fermionic Hubbard model, and on the other hand, may inspire
exact solutions for other models and hence provide  a general method of analyzing exactly solvable fixed points of interacting markovian semigroups. 
The concept of a walking graph state (\ref{eq:wg}) should be explored as a general ansatz for classifying and deriving new solutions in terms of graph diagrams (such as Fig.~\ref{fig:HD}).
For example, the solution of the open $XXZ$ chain of Refs.~\cite{prosen:11a,prosen:11b} can be identified with a {\em linear} semi-infinite chain graph $G$, with nodes ${\cal V}(G)=\{0,1,2\ldots\}$, edges
${\cal E}(G)=\cup_{k=0}^\infty\{(k,k),(k,k+1),(k+1,k)\},$ index function $\omega(k,k)=\sigma^0,\omega(k,k+1)=\sigma^+,\omega(k+1,k)=\sigma^-$, and $\sigma^\z$ as the only possible defect operator.
Furthermore, walking graph state may be implemented as a variational ansatz for efficient numerical simulations of non-integrable models.

Useful remarks from  B. Bu\v ca, E. Ilievski, V. Popkov and B. \v Zunkovi\v c are gratefully acknowledged, as well as support by the grants P1-0044 and J1-5439 of Slovenian Research Agency (ARRS).

\end{document}